\documentclass{llncs}
\usepackage{tabularx,booktabs,multirow,delarray,array}
\usepackage{graphicx,amssymb,amsmath,amssymb}
\usepackage{latexsym}
\usepackage[mathlines]{lineno}

\newcommand{\OPT}{\mbox{$O\!P\!T$}}

\newenvironment{myproof}{\noindent {\textbf{Proof:}}\rm}{\hfill $\Box$\rm}

\def\calP{\mathcal{P}}

\def\calD{\mathcal{D}}
\def\calR{\mathbb{R}^2}
\def\cen{$k$-center}
\def\wsf{WSF}
\def\problem{{\it 1DkCenter}}

\def\sectionspace{\vspace*{-0.10in}}

\begin{document}
%\linenumbers

\title{Efficient Algorithms for the One-Dimensional $k$-Center Problem\thanks{D.Z.~Chen's research was supported in part by NSF
under Grant CCF-1217906. H.~Wang's research was supported in part by NSF under Grant CCF-1317143.}}

\author{Danny Z.~Chen\inst{1} \and Jian Li\inst{2}
\and Haitao Wang\inst{3}\thanks{This work was partially done while the
author was visiting at IIIS at Tsinghua University.}}

\institute{
  Department of Computer Science and Engineering\\
  University of Notre Dame, Notre Dame, IN 46556, USA\\
  \email{dchen@nd.edu}\\
  \and
  Institute for Interdisciplinary Information Sciences (IIIS)\\
  Tsinghua University, Beijing 100084, China\\
  \email{lijian83@mail.tsinghua.edu.cn}\\
  \and
  Department of Computer Science\\
  Utah State University, Logan, UT 84322, USA\\
  \email{haitao.wang@usu.edu}\\
}

\maketitle

\pagestyle{plain}
\pagenumbering{arabic}
\setcounter{page}{1}

\begin{abstract}
We consider the problem of finding $k$ centers for $n$ weighted points on a real line.
This (weighted) \cen\ problem was solved in $O(n\log n)$ time previously by using Cole's parametric search and other complicated approaches. In this paper, we present an easier $O(n\log n)$ time algorithm that avoids the parametric search, and in certain special cases our algorithm solves the problem
in $O(n)$ time. In addition, our techniques involve developing interesting data structures for processing queries that find a
lowest point in the common intersection of a certain subset of half-planes. This subproblem is interesting in its own right and our solution for it may find other applications as well.
\end{abstract}


%\newpage

\section{Introduction}
\label{sec:intro}

We study the weighted $k$-center problem for a set
of $n$ points on a real line. Let $P=\{p_1$, $p_2,\ldots,p_n\}$ be a set
of $n$ points on a real line $L$. For each $i$ with $1\leq i\leq n$,
the point $p_i\in P$ has a weight $w(p_i)\geq 0$. For a point $p$ on
$L$, denote by $L(p)$ the coordinate of $p$ on $L$, which we also
refer to as the $L$-coordinate of $p$. For two points $p$ and $q$ on
$L$, let $d(p,q)=|L(p)-L(q)|$ be the {\em distance} between $p$ and
$q$. Further, for a set $F=\{f_1,f_2,\ldots,f_k\}$ of
points and a point $q$ on $L$, define $d(q,F)=d(F,q)=\min_{1\leq
j\leq k}d(q,f_j)$. Given $P$ and an integer $k>0$, the weighted
{\em one-dimensional $k$-center} problem seeks to determine a set
$F=\{f_1,f_2,\ldots,f_k\}$ of $k$ points on $L$ such that the value
$\psi(P,F)=\max_{p_i\in P}(w(p_i)\cdot d(p_i,F))$ is minimized. We
use \problem\ to denote this problem. Also, the points in $F$ are
called {\em centers}, and the points in $P$ are called
{\em demand points}.
%for each customer $p_i\in P$, the weighted distance
%$d(p_i,F)$ is called the {\em service distance} for $p_i$.

The {\em unweighted version} of \problem\ is the case where all
points have the same weight. If $F\subseteq P$ is required, then the
case is called the {\em discrete version}.

%\sectionspace
%\subsection{Previous Work}

Although many variants of the \cen\ problem are NP-hard
\cite{ref:Ben-MosheAn06,ref:ChandrasekaranPo82,ref:FredericksonFi83,ref:KarivAn79,ref:MegiddoNe83}, some special cases are
solvable in polynomial time. Megiddo and Tamir
\cite{ref:MegiddoNe83} presented an $O(n\log^2 n\log\log n)$ time
algorithm for the weighted \cen\ problem on a tree of $n$ nodes, and
the running time can be reduced to $O(n\log^2 n)$ by applying Cole's
parametric search \cite{ref:ColeSl87}. Later, Frederickson
\cite{ref:FredericksonPa91} gave a linear time algorithm for the
unweighted \cen\ problem on a tree. Jeger and Kariv
\cite{ref:JegerAl85} gave an $O(kn\log n)$ time algorithm for the
weighted \cen\ problem on a tree. For the weighted \cen\ problem on
a real line (i.e., the problem \problem), Bhattacharya and Shi
\cite{ref:BhattacharyaOp07} recently proposed an algorithm with a
time bound linear in $n$ but exponential in $k$.
%In the preliminary of this paper \cite{ref:ChenEfkCenter11},
%an $O(\min\{n\log^{1.5}n,n\log
%n+k^2\log^2{\frac{n}{k}}\log^2n\})$ time algorithm for \problem\ was given.
%Chen and Wang \cite{ref:ChenEfkCenter11} reduce the \problem\ problem to a
%special case of a points approximation problem \cite{ref:ChenAp091}
%and utilize the techniques there.
%Very recently, Fournier and
%Vigneron \cite{ref:FournierA11} announced a new $O(n\log n)$ time
%algorithm for the above points approximation problem based on the
%parametric search in \cite{ref:ColeSl87}, Thus, by combining the
%results in \cite{ref:ChenEfkCenter11} and \cite{ref:FournierA11}, the
%\problem\ problem can be solved in $O(n\log n)$ time.
In addition, the discrete weighted \cen\ problem on a tree is solvable in
$O(n\log^2 n)$ time \cite{ref:MegiddoAn81} and the discrete
unweighted \cen\ problem on a tree is solvable in $O(n)$ time
\cite{ref:FredericksonPa91}.
The discrete weighted \problem\ has been solved in
$O(n\log n)$ time \cite{ref:MegiddoAn81}, without using the parametric search in \cite{ref:ColeSl87}.
Note that our problem \problem\ is the
``non-discrete'' weighted version.

As indicated by Tamir \cite{ref:TamirPe14},
 the problem \problem\ can be solved in $O(n\log n)$ time by combining the techniques given by Megiddo and Tamir \cite{ref:MegiddoNe83} and Cole's parametric search  \cite{ref:ColeSl87}.
Although the approach in \cite{ref:MegiddoNe83} is elegant and simple, Cole's parametric search \cite{ref:ColeSl87} is quite complicated and involves large constants, and thus is mainly of theoretical interest.
In this paper, we present another $O(n\log n)$ time algorithm for \problem, which is much easier and avoids Cole's parametric search \cite{ref:ColeSl87}.
Further, if all points in $P$ are given sorted on $L$ and their weights are
also sorted, our algorithm can solve \problem\ in
$O(n+k^2\log^2\frac{n}{k}\log n\log\log n)$ time, which is in favor
of small $k$.
For example, if $k=O(n^{1/2-\epsilon})$ for any $\epsilon>0$ (which is true
in many applications), our algorithm runs in $O(n)$ time.
It should be noted that if the points in $P$ are given sorted on $L$, then the unweighted \problem\ is solvable in $O(k^2\log^2 n)$ time \cite{ref:TamirPe14}.
In addition, our techniques also yield an efficient data structure for
processing queries for finding a lowest point in the common
intersection of a certain subset of half-planes, which we call the
{\em 2-D sublist LP queries}. Since the 2-D sublist LP query is a basic
geometric problem, our data structure may be interesting in its own right.

\subsection{An Overview of Our Approach}

We first model the \problem\ problem as a problem of approximating a set of
weighted points by a step function in the plane \cite{ref:ChenA13,ref:ChenAp13,ref:FournierA13,ref:LiuA10}.
For this points approximation problem, by using Cole's parametric search \cite{ref:ColeSl87}, Fournier and
Vigneron \cite{ref:FournierA13} gave an $O(n\log n)$ time
algorithm. Thus, by combining
our problem modeling and the algorithm in \cite{ref:FournierA13}, the
\problem\ problem can be solved in $O(n\log n)$ time. This approach, again, uses Cole's parametric search \cite{ref:ColeSl87}. Chen and Wang \cite{ref:ChenA13} proposed another $O(n\log n)$ time algorithm without using parametric search, by modifying the slope selection algorithms \cite{ref:BronnimannOp98,ref:KatzOp93}. Although the algorithm in \cite{ref:ChenA13} avoids the parametric search, it is still complicated and not practical because it involves the techniques of either cutting \cite{ref:BronnimannOp98} or expanders \cite{ref:KatzOp93}. In addition, Liu \cite{ref:LiuA10} presented an $O(n\log n)$ time randomized algorithm for this points approximation problem.

%Two algorithms were given
%in \cite{ref:ChenAp091} for this point approximation problem with the time bounds of
%$O(n\log^2 n)$ and $O(n\log n+
%k^2\log^2\frac{n}{k}\log^2 n)$, respectively. Consequently, the
%\problem\ problem can be solved in $O(\min\{n\log^2 n,n\log n+
%k^2\log^2\frac{n}{k}\log^2 n\})$ time.

In fact, we model \problem\ as a special case of the above points approximation problem such that we are able to develop a simple and deterministic $O(n\log n)$ time algorithm. Further, we can solve the problem in $O(n)$ time in some special situations, as discussed earlier.
%However, the \problem\ problem has some special properties that
%allow us to develop faster solutions.
Specifically, after the
geometric transformations, a key component to solving the problem is
the following {\em 2-D
sublist LP query} problem: Given a set of $n$ upper half-planes,
$H=\{h_1,h_2,\ldots,h_n\}$, in the plane, for each query $q(i,j)$
($1\leq i\leq j\leq n$), compute a lowest point $p^*$ in the common
intersection of all half-planes in $H_{ij}=\{h_t \ | \ i\leq t\leq
j\}$. A data structure was proposed in \cite{ref:ChenAp091} for this
problem, which can be built in $O(n\log n)$ time and answers each
query in $O(\log^2 n)$ time.
On the \problem\ problem, we observe that the input half-plane
set $H$ has a special property that the intersections between the
$x$-axis and the bounding lines of the half-planes are
ordered from left to right according to the half-plane indices in $H$. Exploiting
this special property and using the compact
interval trees \cite{ref:GuibasCo91},
we design a new data structure for this special case of the 2-D
sublist LP queries, which can be built in
$O(n\log n)$ time and can answer each query in $O(\log n)$
time. This new data structure allows us to solve \problem\ in
$O(n\log n)$ time, and in $O(n)$ time in certain situations.
Further, since the 2-D sublist LP query problem
is a very basic problem, our new data structure may find other applications as
well.

%We should mention that very recently, by using Cole's
%parametric search \cite{ref:ColeSl87}, Fournier and
%Vigneron \cite{ref:FournierA13} gave an $O(n\log n)$ time
%algorithm for the above point approximation problem \cite{ref:ChenAp091}.
%Thus, by combining
%our problem modeling and the algorithm in \cite{ref:FournierA13}, the
%\problem\ problem can be solved in $O(n\log n)$ time. However, as pointed out in \cite{ref:FournierA13}, the parametric search approach in \cite{ref:ColeSl87} is
%quite complicated and involves large constants, and thus the algorithm in \cite{ref:FournierA13} is mainly of theoretical interest.
%In contrast, our approach is much simpler and more practical.

In the following, we present the high-level scheme of our algorithm
in Section \ref{sec:algorithms}. In Section
\ref{sec:chreduction}, we model our problem as the
2-D sublist LP queries and present our data structure.
Section \ref{sec:conclusion} concludes the
paper and discusses the $\Omega(n\log n)$ time lower bound of the problem \problem.

For simplicity of discussion, we make a general position assumption that no two
points in $P$ are at the same position on $L$. We also assume the
weight of each point in $P$ is positive and finite. These
assumptions are only for ease of exposition and our algorithms can
be easily extended to the general case.

\section{The Algorithmic Scheme}
\label{sec:algorithms}

In this section, we discuss the high-level framework
of our algorithm. As pointed out in
\cite{ref:BhattacharyaOp07}, it is possible that there is more than
one optimal solution for the \problem\ problem.
Our algorithm focuses on finding one optimal solution.

\subsection{Preliminaries}

For any two points $p$ and $q$ on $L$ with $L(p)\leq L(q)$ (recall
that $L(p)$ is the coordinate of $p$ on L, and similarly for $L(q)$), denote
by $[p,q]$ the (closed) interval of $L$ between $p$ and $q$.

We first sort all points of $P$ from left to right on $L$. Without loss of
generality (WLOG), let $\{p_1, p_2,\ldots,p_n\}$ be the sorted order in
increasing coordinates on $L$. For any two points $p_i,p_j \in
P$ with $i\leq j$, denote by $I(i,j)$ the interval $[p_i,p_j]$. Let
$\psi^*$ be the value of $\psi(P,F)$ for an optimal solution of
\problem. Suppose $F$ is the center set in an optimal solution;
for a demand point $p\in P$
and a center $f\in F$, if $(w(p)\cdot d(f,p))\leq \psi^*$, then we say that
$p$ can be {\em served} by $f$.  It is easy to see that
there is an optimal solution $F$ such that each center of $F$ is in
$[p_1,p_n]$. Further, as discussed in \cite{ref:BhattacharyaOp07},
there is an optimal solution $F$ such that the
points of $P$ are partitioned into $k$ intervals $I(1,i_1),
I(i_1+1,i_2), \ldots, I(i_{k-1}+1,n)$ by integers $i_0+1=1\leq i_1\leq
i_2\leq \cdots\leq i_{k-1}\leq n =i_k$, each interval $I(i_{j-1}+1,i_j)$
contains exactly one center $f_j\in F$, and for each point $p\in
P\cap I(i_{j-1}+1,i_j)$, $(w(p)\cdot d(f_j,p))\leq \psi^*$
holds. In other words, each center of
$F$ serves a subset of consecutive demand points in $P$.

For any two integers $i$ and $j$ with $1\leq i\leq j\leq n$, denote
by $P_{ij}$ the subset of points of $P$ in the interval $I(i,j)$, i.e.,
$P_{ij}=\{p_i,p_{i+1},\ldots,p_j\}$ ($P_{ij}=\{p_i\}$ for $i=j$).
Consider the following {\em weighted 1-center} problem: Find a
single center (i.e., a point) $f$ in the interval $I[i,j]$ such that
the value of $\psi(P_{ij},f)=\max_{p_t\in P_{ij}}(w(p_t)\cdot d(p_t,f))$ is
minimized. Let $\alpha(i,j)$ denote the minimum value of
$\psi(P_{ij},f)$ for this weighted 1-center problem.

For solving the \problem\ problem, our strategy is to determine $k-1$
integers $1\leq i_1\leq i_2\leq \cdots\leq i_{k-1}\leq n$ such that the value of
$\max\{\alpha(1,i_1),\alpha(i_1+1,i_2),\ldots, \alpha(i_{k-1}+1,n)\}$
is minimized and this minimized value is $\psi^*$. Note that
in the above formulation, for each value $\alpha(i,j)$, exact one center is
determined in the interval $I(i,j)$. To solve this problem, we reduce
it to a planar weighted point approximation problem
\cite{ref:ChenAp091} in the next subsection.

\subsection{The Reduction to the Planar Weighted Point Approximation Problem}
\label{sec:reduction}

We first review the planar weighted point approximation problem
and then show our problem reduction.

\begin{figure}[t]
\begin{minipage}[t]{\linewidth}
\begin{center}
\includegraphics[totalheight=0.8in]{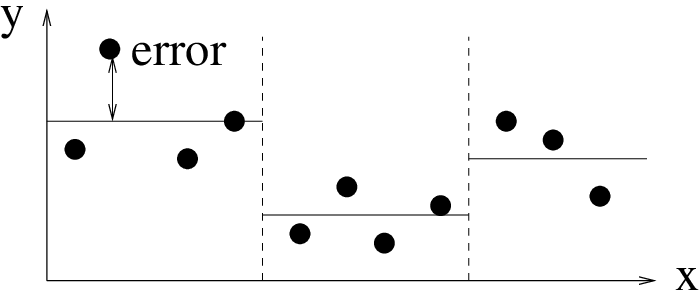}
\caption{\footnotesize Approximating a set of points by a step function
(i.e., the three horizontal line segments).}
\label{fig:sfpf}
\end{center}
\end{minipage}
\vspace*{-0.15in}
\end{figure}

Let $P'=\{p'_1, p'_2, \ldots, p'_n\}$ be a point set in the plane
with $p'_i=(x_i,y_i)$, and each point $p'_i$ be associated with a
weight $w(p_i')\geq 0$. Assume the points in $P'$ are ordered
increasingly by their $x$-coordinates. Suppose $g$ is a step
function (i.e., a peicewise constant function, e.g., see Fig.~\ref{fig:sfpf})
which we use to approximate the points of $P'$ (in other
words, we fit the step function $g$ to the point set $P'$). The {\em
weighted vertical distance} between any point $p'_i\in P'$ and $g$ is
defined as $d_w(p_i',g)=w(p_i')\cdot |y_i-g(x_i)|$ (see
Fig.~\ref{fig:sfpf}). The approximation error of $g$, denoted by
$e(P',g)$, is defined as $\max_{p'_i\in P'} d_w(p'_i,g)$. The {\em
size} of $g$ is the number of its horizontal line segments. Given an
integer $k>0$, the {\em point approximation problem} seeks a step
function $g$ to approximate the points of $P'$ such that the size of
$g$ is at most $k$ and the error $e(P,g)$ is minimized. In
\cite{ref:ChenAp091}, this problem is referred to as the {\em
weighted step function min-$\epsilon$ problem}, denoted by \wsf.
Here we also use \wsf\ to denote this problem.

We now show that the \problem\ problem can be reduced to \wsf.
Consequently, \wsf\ algorithms can be used to solve \problem.
Indeed, consider the demand point set $P=\{p_1,p_2,\ldots,p_n\}$ for
\problem\ with the points ordered increasingly
by their coordinates on $L$. For each demand point $p_i\in P$,
$1\leq i\leq n$, we create a point $p'_i=(i,L(p_i))$ in a 2-D
Euclidean plane $\calR$
(i.e, the $x$-coordinate of $p'_i$ in $\calR$ is the index $i$ and
its $y$-coordinate $y'_i$ in $\calR$ is the coordinate of $p_i$ on
$L$), and let the weight of $p'_i$ be that of $p_i$ (i.e.,
$w(p'_i)=w(p_i)$). Let $P'$ be the set of $n$ weighted points thus
created in $\calR$. The next lemma states the relation between
\problem\ and the reduced instance of \wsf.

\begin{lemma}\label{lem:00}
An optimal solution $\OPT_{P'}$ for \wsf\ on $P'$ in $\calR$
corresponds to an optimal solution $\OPT_P$ for \problem\ on $P$.
Further, once having $\OPT_{P'}$, $\OPT_P$ can be obtained in $O(n)$
time.
\end{lemma}

\begin{myproof}
For any two integers $i$ and $j$ with $1\leq i\leq j\leq n$, let
$P'_{ij}=\{p'_i,p'_{i+1},\ldots,p'_j\}$ ($P'_{ij}=\{p'_i\}$ for
$i=j$). Consider the following problem: Find a value $Y$ for one
single horizontal line segment with $Y$ as its $y$-coordinate such
that the value of $d_w(P'_{ij},Y)=\max_{p'_t\in
P'_{ij}}(w(p'_t)\cdot |y_t-Y|)$ is minimized (where $y_t=L(p_t)$).
Let $\alpha'(i,j)$ denote the minimized value of $d_w(P'_{ij},Y)$.

Let $\epsilon^*$ be the approximation error of an optimal solution
for \wsf\ on $P'$. It is easy to see that computing an optimal
solution for \wsf\ on $P'$ is equivalent to determining $k-1$
integers $1\leq i_1\leq i_2\leq \cdots\leq i_{k-1}\leq n$ such that
the value of $\max\{\alpha'(1,i_1),\alpha'(i_1+1,i_2),\ldots,
\alpha'(i_{k-1}+1,n)\}$ is minimized and this minimized value is
$\epsilon^*$.
According to the way that we create the point set $P'$ from the
demand point set $P$, each value $\alpha'(i,j)$ is exactly equal to
the value $\alpha(i,j)$, which is the minimized value of
$\psi(P_{ij},f)$ for the weighted 1-center problem on the demand
point subset $P_{ij}$ by determining the value of $f$.
Further, we have shown that to find an optimal solution for
\problem\ on $P$, it suffices to determine $k-1$ integers $1\leq
i_1\leq i_2\leq \cdots\leq i_{k-1}\leq n$ such that the value of
$\max\{\alpha(1,i_1),\alpha(i_1+1,i_2),\ldots,
\alpha(i_{k-1}+1,n)\}$ is minimized and the minimized value is
$\psi^*$.

The above discussion shows that to find an optimal solution for
\problem\ on $P$, it suffices to find an optimal solution for \wsf\
on $P'$; further, $\psi^*=\epsilon^*$. Given an optimal solution
$\OPT_{P'}$ for \wsf\ on $P'$, below we show how to obtain an
optimal solution $\OPT_P$ for \problem\ on $P$ from $OPT_{P'}$ in linear time.

Note that $\OPT_{P'}$ is a step function with $k$ steps (i.e.,
horizontal line segments).  Let $i_0=0$ and $i_k=n$. For each $1\leq
j\leq k$, suppose the $j$-th step of $\OPT_{P'}$ has a
$y$-coordinate $y^j$ and covers the points of $P'$ from
$p'_{i_{j-1}+1}$ to $p'_{i_j}$, i.e., for each point $p_t'\in P'$
with $i_{j-1}+1\leq t\leq i_j$, the vertical line through $p_t'$
intersects the $j$-th horizontal segment of $\OPT_{P'}$.
We obtain $\OPT_P$ for \problem\ on $P$ as follows. For each $1\leq
j\leq k$, the $j$-th center $f_j$ is put at the position $y^j$ on
$L$ (i.e., $L(f_j)=y^j$), which serves the demand points of $P$ from
$p_{i_{j-1}+1}$ to $p_{i_j}$.

Thus, once $\OPT_{P'}$ is available, $\OPT_P$ can be obtained in
$O(n)$ time.
\end{myproof}

Based on Lemma \ref{lem:00}, to compute a set $F$ of $k$ centers for
$P$ to minimize the value $\psi(P,F)$, it suffices to solve the
corresponding WSF problem on $P'$ and $k$. Specifically, after an
optimal step function $g$ for $P'$ is obtained, each horizontal
segment of $g$ defines a center on $L$ whose coordinate is equal to
the $y$-coordinate of that horizontal segment of $g$ in $\calR$.

To apply the WSF algorithms to the \problem\ problem, we need a data
structure for answering queries $q(i,j)=\alpha(i,j)$ with $1\leq i\leq
j\leq n$. Suppose such a data structure can be built in $O(\pi(n))$
time and can answer each query $\alpha(i,j)$ in $O(q(n))$ time; then we
say the time bounds of the data structure are $O(\pi(n),q(n))$. The two
lemmas below follow from the results in \cite{ref:ChenAp091}.

%The algorithm is based on Theorem $2$ in \cite{ref:FournierFi08}
%which is a general formulation of the path partition algorithm
%\cite{ref:FredericksonOp91}. We rewrite the theorem in the following
%lemma.
%
%\begin{lemma}\cite{ref:FournierFi08}
%Suppose $\theta(i,j)$ is a function where $i$ and $j$ are two
%integers with $1\leq i\leq j\leq n$ such that $\theta(i,j)\geq 0$
%and $\theta(i,j)=0$ when $i=j$. Define the problem MIN-MAX
%PARTITION($\theta$) as follows: Partition interval $(1,2,\ldots,n)$
%into $k$ subintervals $I_1,\ldots,I_k$ where each subinterval $I_i$
%consists of consecutive integers from $l_{i-1}+1$ to $l_i$ (suppose
%$l_0=1$ and $l_k=n$), such that $\max_{1\leq i\leq
%k}\theta(l_{i-1}+1,l_i)$ is minimized. If $\theta$ has the following
%properties:
%\begin{enumerate}
%\item
%$\theta$ is non-decreasing, that is, $\theta(i,j)\leq \theta(i',j')$
%for $1\leq i'\leq i\leq j\leq j'\leq n$. \vspace*{-0.10in}
%\item
%With $\pi(n)$ time preprocessing, for any integer pairs $i\leq j$,
%$\theta(i,j)$ can be computed in $q(n)$ time.
%\end{enumerate}
%
%Then MIN-MAX PARTITION($\theta$) problem can be solved in
%$(\pi(n)+n\cdot q(n))$ time.
%\end{lemma}
%
%For our problem, if we let $\theta(i,j)=\alpha(i,j)$, it is easy to
%see that $\theta(i,j)=0$ when $i=j$ and $\theta$ is non-decreasing.
%Therefore we have the following lemma.

\begin{lemma}\label{lem:10}{\em \cite{ref:ChenAp091}} Suppose
there is a data structure for the queries $\alpha(i,j)$ with time
bounds $O(\pi(n),q(n))$; then the \problem\ problem is solvable
in $O(\pi(n)+n\cdot q(n))$ time.
\end{lemma}

\begin{lemma}\label{lem:20}{\em \cite{ref:ChenAp091}} Suppose
there is a data structure for the queries $\alpha(i,j)$ with time
bounds $O(\pi(n),q(n))$; then the \problem\ problem is solvable
in $O(\pi(n)+q(n)\cdot k^2\log^2\frac{n}{k})$ time.
\end{lemma}

Refer to \cite{ref:ChenAp091} for the details of the algorithms in the
above two lemmas.
By the above two lemmas, Lemma \ref{lem:100} follows.

\begin{lemma}\label{lem:100} Suppose
there is a data structure for the queries $\alpha(i,j)$ with time
bounds $O(\pi(n),q(n))$; then the \problem\ problem can be solved
in $O(\min\{\pi(n)+n\cdot q(n),\pi(n)+q(n)\cdot k^2\log^2\frac{n}{k}\})$ time.
\end{lemma}

%Using such data structures, the algorithm for Lemma \ref{lem:10} follows an approach in
%\cite{ref:FournierFi08}, which is a general formulation of the path
%partition scheme \cite{ref:FredericksonOp91}; the algorithm for
%Lemma \ref{lem:20} is based on parametric search \cite{ref:ColeSl87,ref:MegiddoAp83}
%and exponential search (refer to \cite{ref:ChenAp091} for the
%details of these algorithms).

A data structure based on fractional cascading \cite{ref:ChazelleFr86} was given in
\cite{ref:ChenAp091} for answering the
queries $\alpha(i,j)$ with time bounds $O(n\log n,\log^2 n)$. Consequently, by
Lemma \ref{lem:100}, the \problem\ problem is
solvable in $O(\min\{n\log^2 n, n\log n+ k^2\log^2\frac{n}{k}\log^2
n\})$ time. In Section \ref{sec:chreduction}, we develop a data
structure for processing the queries $\alpha(i,j)$ with time bounds
$O(n\log n, \log n)$, which allows us to solve \problem\
in $O(n\log n)$ time.

The reason why we can solve the \problem\ problem faster than simply applying
the \wsf\ algorithms \cite{ref:ChenAp091} is that the
\wsf\ instance constructed above from the problem \problem\ has a special property: The
$y$-coordinates of the points $p_1',p_2',\ldots,p_n'$ are increasing.
As shown in Section \ref{sec:chreduction}, this
special property allows us to design a new data structure for the $\alpha(i,j)$
queries with time bounds $O(n\log n, \log n)$.
Note that this special property does not hold for the general \wsf\
problem studied in \cite{ref:ChenAp091}.

\section{The Data Structure for Computing $\alpha(i,j)$}
\label{sec:chreduction}

In this section, we present a data structure with time bounds
$O(n\log n, \log n)$ for answering the $\alpha(i,j)$ queries.
In the following, we first model the problem of computing $\alpha(i,j)$
as the problem of finding a lowest point in the common
intersection of a set of half-planes (i.e., the 2-D sublist LP query).

\sectionspace
\subsection{The Problem Modeling}

Consider a point subset $P_{ij}\subseteq P$ with $i\leq j$. Recall that
$L(p_i)\leq L(p_{i+1})\leq\cdots\leq L(p_j)$. To compute $\alpha(i,j)$, we
need to find a point $f$ such that the value $\psi(P_{ij},f)=
\max_{i\leq t\leq j}(w(p_t)\cdot
d(f,p_t))=\max_{i\leq t\leq j}(w(p_t)\cdot|L(f)-L(p_t)|)$ is minimized and
$\alpha(i,j)$ is the minimized value.
Consider an arbitrary point $f'$ on $L$.
Since $\psi(P_{ij},f') =
\max_{i\leq t\leq j}(w(p_t)\cdot|L(f')-L(p_t)|)$, each point
$p_t\in P_{ij}$ defines two constraints:
$w(p_t)\cdot (L(f')-L(p_t))\leq \psi(P_{ij},f')$ and
$-w(p_t)\cdot (L(f')-L(p_t))\leq \psi(P_{ij},f')$.

Consider a 2-D $xy$-coordinate system with $L$ as the $x$-axis. For each
point $p_t\in P$, the inequality $w(p_t)\cdot|x-L(p_t)|\leq y$ defines
two (upper) half-planes: $w(p_t)\cdot (x-L(p_t))\leq y$ and $-w(p_t)\cdot
(x-L(p_t))\leq y$.  Note that the two lines bounding the two half-planes
intersect at the point $p_t$ on $L$.

\begin{figure}[t]
\begin{minipage}[t]{\linewidth}
\begin{center}
\includegraphics[totalheight=1.1in]{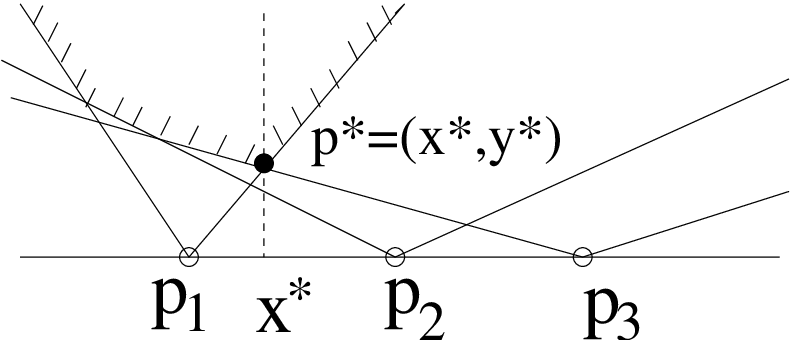}
\caption{\footnotesize Illustrating the common intersection of the
half-planes defined by three points $p_1$, $p_2$, and $p_3$. The
point $p^*$ is the lowest point in the common
intersection.}\label{fig:cone}
\end{center}
\end{minipage}
\vspace*{-0.15in}
\end{figure}

Based on the above discussion, if $p^*=(x^*,y^*)$ is a lowest point in
the common intersection of the $2(i-j+1)$ (upper) half-planes defined by the
points in $P_{ij}$, then $\alpha(i,j)=y^*$ and $L(f)=x^*$ is the
coordinate of an optimal center $f$ on $L$ for $P_{ij}$. Figure~\ref{fig:cone}
shows an example in which each ``cone" is the intersection of
the two upper half-planes defined by a point in $P_{ij}$.
%Thus, to compute $\alpha(i,j)$, it suffices to compute the lowest point $p^*$.
Clearly, this is an
instance of the 2-D linear programming (LP) problem, which is solvable
in $O(j-i+1)$ time \cite{ref:DyerLi84,ref:MegiddoLi84}. However, we
can make the computation faster by preprocessing. Let
$H_P=\{h_1,h_2,\ldots,h_{2n}\}$ be the set of $2n$ (upper) half-planes defined by
the $n$ points in $P$, such that for each $1\leq i\leq n$, the demand point
$p_i$ defines $h_{2i-1}$ and $h_{2i}$. Then to compute
$\alpha(i,j)$, it suffices to find the lowest point $p^*$ in the
common intersection of the half-planes defined by the points in
$P_{ij}$, i.e., the half-planes in
$H_{2i-1,2j}=\{h_{2i-1},h_{2i},h_{2i+1},h_{2i+2},\ldots,h_{2j-1},h_{2j}\}$.

We actually consider a more general problem: Given in the plane
a set of $n$ upper half-planes $H=\{h_1,h_2,\ldots,h_n\}$, each
query $q(i,j)$ with $1\leq i\leq j\leq n$ asks for a lowest point
$p^*$ in the common intersection of all half-planes in
$H_{ij}=\{h_t \ | \ i\leq t\leq j\}\subseteq H$. We call this problem the {\em
2-D sublist LP query}. Based on our discussion above,
if we solve the 2-D sublist LP query problem, then the
$\alpha(i,j)$ queries for the \problem\ problem can be processed as well in the
same time bound.
%The 2-D sublist LP query problem has been studied before, and
A data structure for the 2-D sublist LP query problem with time bounds $O(n\log n,
\log^2 n)$ was given in \cite{ref:ChenAp091}.

Yet, the 2-D sublist LP query problem for
\problem\ is special in the
following sense. For each half-plane $h\in H_P$ for \problem, we call the
$x$-coordinate of the intersection point between $L$ (i.e., the $x$-axis)
and the line bounding $h$ the {\em $x$-intercept} of $h$
(or its bounding line). As discussed above, for each
point $p_t\in P$, the $x$-intercepts of both the half-planes $h_{2t-1}$
and $h_{2t}$ defined by $p_t$ are exactly the point $p_t$. Since all
points of $P$ are ordered along $L$ from left to right by their
indices, a special property of $H_P$ is that the $x$-intercepts of
all half-planes in $H_P$ are ordered from left to right on $L$ by
the indices of the half-planes. For a set $H$ of half-planes for a
2-D sublist LP query problem instance, if $H$ has the above special property,
then we say that $H$ is {\em $x$-intercept ordered}.
%Note that this special property of $H_P$ is due to the
%special property of the \wsf\ problem instance to which the \problem\ problem is
%reduced (i.e., all points in $P'$ are ordered increasingly in
%their $y$-coordinates from $p_1'$ to $p_n'$), as discussed in Section
%\ref{sec:reduction}.

Below, we show that if $H$ is $x$-intercept ordered, then there is a
data structure for the specific 2-D sublist LP query problem with time
bounds $O(n\log n, \log n)$.
%In the following discussion,
Henceforth, we assume that $H$ is an $x$-intercept ordered half-plane
set. In the \problem\ problem, since all the point weights for $P$
are positive finite values, the bounding line of each half-plane in
$H_P$ is neither horizontal nor vertical. Thus, we also assume that
no bounding line of any half-plane in $H$ is horizontal
or vertical. Again, this assumption is only for simplicity of discussion.
% and our result can be generalized to the general case.

In the next section, we solve the 2-D sublist LP queries by reducing
it to computing the convex hull of a query sub-path of a given simple
path \cite{ref:GuibasCo91}.
Given a simple path in the plane, based on
compact interval trees, data structures are proposed in
\cite{ref:GuibasCo91} to compute (in logarithmic time)
the convex hull of a query subpath
that is specified by the indices of the beginning vertex and the end
vertex of the subpath, and the convex hull is represented
(by a compact interval tree) such that standard convex hull
queries on it can be done in $O(\log n)$ time.

\subsection{Answering 2-D Sublist LP Queries}
%\label{sec:chreduction}

%In this section, we give a data structure of time complexity
%$O(n\log n,\log n)$ for the 2-D sublist LP queries, which is done by
%reducing the 2-D sublist LP queries to a subpath convex hull queries of a simple
%path \cite{ref:GuibasCo91}. If the slopes of the bounding lines of
%the half-planes in $H$ are sorted, then more efficient data structures
%are also presented.  For simplicity of discussion, we assume
%the bounding line of any upper half-plane in $H$ is not vertical or
%horizontal and
%the bounding lines of any two upper-half-planes in $H$ do not
%intersect the $x$-axis at the same point. Note that these assumptions
%are only for ease of exposition and our data structures can be easily
%generalized to handle the general situation.
For each half-plane $h_i\in H$, we denote by $l(h_i)$ the
bounding line of $h_i$; let $l(H')$ be the set of the bounding lines of
the half-planes in any subset $H'\subseteq H$.

Our problem reduction utilizes a duality transformation
\cite{ref:deBergCo08}, which is a technique commonly used in
computational geometry, as follows. Suppose we have a
{\em primary} plane $\calP$. For each point $(a,b)\in\calP$, it
corresponds to a line $y=ax-b$ in the {\em dual} plane $\calD$; the
line is also called the {\em dual} of the point and vice versa.
Similarly, each line $y=a'x-b'$ in $\calP$ corresponds to a point
$(a',b')$ in $\calD$.

Suppose all half-planes in $H$ are in the primary plane $\calP$.
By duality, we can obtain a
set $H^*$ of points in the dual plane $\calD$ corresponding to the
lines in $l(H)$. For each query $q(i,j)$ on $H$, our goal is to locate
the lowest point $p^*$ in the common intersection of all half-planes
in $H_{ij}$ (note that due to our assumption that no line in $l(H)$ is
horizontal, there is only one lowest point in the common intersection).
Since all half-planes in $H$ are upper half-planes,
an observation is that $p^*$ is also the lowest point of
the upper envelope of the arrangement of the lines in $l(H_{ij})$.
Denote by $U_{ij}$ the above upper envelope for $l(H_{ij})$.
Denote by $H^*_{ij}$ the set of points (in the dual plane $\calD$) dual to the
lines  in $l(H_{ij})$. Let $C^*_{ij}$ denote
the lower hull of the convex hull of $H^*_{ij}$. It is commonly known
\cite{ref:deBergCo08} that the dual of $U_{ij}$ is exactly $C^*_{ij}$
(in the dual plane $\calD$).
Therefore, if we have a representation of $C^*_{ij}$
that can support standard binary-search-based queries, then we can compute
the lowest point $p^*$ in logarithmic time accordingly.

One may attempt to design a data structure for querying
the lower hull on any subset $H^*_{ij}$ of $H^*$. However,
there are difficulties when doing so ``directly", which will be explained later.
Instead, we use an ``indirect" approach, as follows.
Recall that we have assumed the bounding line of any half-plane in
$H$ is not vertical or horizontal.  We partition the half-plane set $H$ into
two subsets $H_1$ and $H_2$ such that a half-plane $h_i$ of $H$ is in $H_1$ (resp.,
$H_2$) if and only if the slope of $l(h_i)$
is negative (resp., positive). Accordingly, for each
subset $H_{ij}$, we also have $H^1_{ij}$ and $H^2_{ij}$,
and we define the envelopes $U^1_{ij}$ and $U^2_{ij}$ accordingly. Since the
bounding lines of all half-planes in $H^1_{ij}$ have negative slopes,
the upper envelope $U^1_{ij}$ is monotone decreasing from left to
right. Similarly,  the upper envelope $U^2_{ij}$ is monotone
increasing from left to right. Hence, it is easy to see that the
lowest point $p^*$ is the {\em single} intersection of $U^1_{ij}$ and $U^2_{ij}$.
Let $H^{*1}_{ij}$ and $H^{*2}_{ij}$ be the sets of points in the dual plane
corresponding to the lines in $l(H^1_{ij})$ and $l(H^2_{ij})$, respectively.
Denote by $C^{*1}_{ij}$ and $C^{*2}_{ij}$ the lower hulls of
$H^{*1}_{ij}$ and $H^{*2}_{ij}$, respectively. By duality, the
intersection of $U^1_{ij}$ and $U^2_{ij}$ corresponds exactly to the
common tangent line of the two lower hulls $C^{*1}_{ij}$ and
$C^{*2}_{ij}$ such that both hulls are above the tangent line. Note
that since all lines in $H^{1}_{ij}$ have negative slopes, by
duality, all points in $H^{*1}_{ij}$ are to the left of the $y$-axis
in the dual plane, and thus the lower hull $C^{*1}_{ij}$ is to the
left of the $y$-axis. Similarly, the lower hull $C^{*2}_{ij}$ is to
the right of the $y$-axis. Namely, the two lower hulls $C^{*1}_{ij}$
and $C^{*2}_{ij}$ are on different sides of the $y$-axis. This
property can make our computation of their tangent line easier.

In summary,
if we can represent both $C^{*1}_{ij}$ and $C^{*2}_{ij}$ in such a way that
the common tangent line can be found efficiently, then $p^*$
can be obtained immediately. Our remaining task is to derive a way to
support convex hull (or lower hull) queries on any subset of consecutive points in
$H^*_1$ (and similarly
on $H^*_2$). Our result is that a data structure can be built in
$O(n\log n)$ time such that given any $i\leq j$, the lower hull
$C^{*1}_{ij}$ can be obtained (implicitly) in $O(\log n)$ time and is represented
in a way that supports binary-search-based queries (e.g., compute the
common tangent of it and another lower hull, say $C^{*2}_{ij}$). The
details are given below.

Without loss of generality, we assume the
bounding lines of the half-planes in $H$ all have negative slopes
(i.e., $H=H_1$) and the other case can be handled analogously. Suppose
the bounding line of each $h_i\in H$ corresponds to the point $h_i^*\in H^*$
in the dual plane $\calD$. Let $\gamma$ be the path by connecting all
pairs of two consecutive points in $H^*$ by line segments, i.e.,
connecting $h^*_i$ to $h^*_{i+1}$
for $i=1,2,\ldots,n-1$. Consider the line segment connecting $h^*_i$ and
$h^*_{i+1}$ and the line segment connecting $h^*_j$ and $h^*_{j+1}$; then
the two segments are {\em adjacent} to each other if $i+1=j$ or
$j+1=i$. The following Lemma \ref{lem:simple} shows that the path
$\gamma$ is a simple path, that is, any two line segments of $\gamma$
that are not adjacent do not intersect.
As can be seen from the proof of Lemma \ref{lem:simple}, we note that
the correctness of Lemma \ref{lem:simple} heavily relies on two properties of $H$:
(1) $H$ is $x$-intercept ordered; (2) the slopes of the bounding lines
of the half-planes in $H$ are all negative (or positive). Without
either property above, the lemma would not hold, and the second
property also
explains why we need to partition the original set $H$ into $H_1$ and
$H_2$.

\begin{lemma}\label{lem:simple}
The path $\gamma$ is a simple path.
\end{lemma}
\begin{myproof}
Consider two segments $s_i$ and $s_j$ where $s_i$ connects $h^*_i$
and $h^*_{i+1}$ and $s_j$ connects $h^*_j$ and $h^*_{j+1}$. Suppose
$s_i$ and $s_j$ are not adjacent. To prove the lemma, it is sufficient
to show $s_i$ and $s_j$ does not intersect.
Since $s_i$ and $s_j$ are not adjacent, either $i+1<j$ or $j+1<i$.
Without loss of generality, assume $i+1<j$.

Note that $s_i$ and $s_j$ are in the dual plane $\calD$.
It is commonly known \cite{ref:deBergCo08} that
the dual of $s_i$ in the primary plane $\calP$ is the {\em double
wedge} bounded by the lines $l(h_i)$ and $l(h_{i+1})$ such that
the double wedge does not contain a vertical line
(e.g., see Fig.~\ref{fig:doublewedge}); we denote the double wedge by $dw(s_i)$.
Similarly, the dual of $s_j$ is the double wedge $dw(s_j)$ bounded by
$l(h_j)$ and $l(h_{j+1})$.

\begin{figure}[t]
\begin{minipage}[t]{\linewidth}
\begin{center}
\includegraphics[totalheight=1.3in]{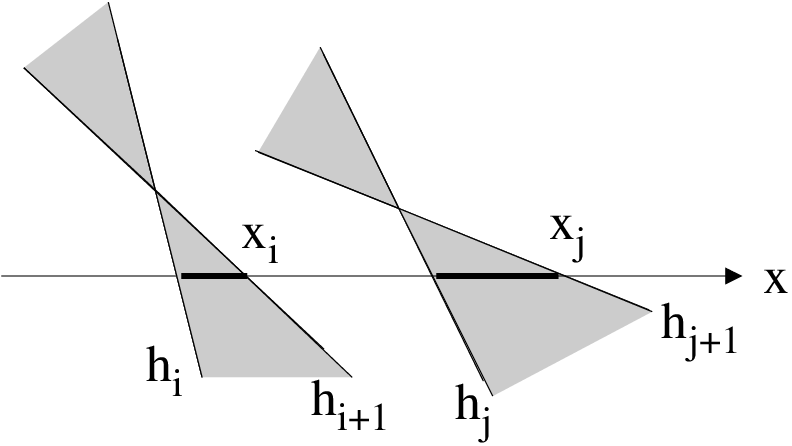}
\caption{\footnotesize Illustrating the two double edges $dw(s_i)$ and
$dw(s_j)$ (the shaded
regions) in the primary plane $\calP$ corresponding to the two segments
$s_i$ and $s_j$ in the dual plane $\calD$. The two segments $x_i$ and
$x_j$ are their intersections with the $x$-axis.}\label{fig:doublewedge}
\end{center}
\end{minipage}
\vspace*{-0.15in}
\end{figure}

Assume to the contrary the two segments $s_i$ and $s_j$ intersect each
other, say, at a point $p$.
Then, $p$ corresponds to a line in the primary
plane $\calP$, which
is in the common intersection of the two double wedges $dw(s_i)$
and $dw(s_j)$. However, we claim that the
common intersection of $dw(s_i)$ and $dw(s_j)$ does not contain any
line, which incurs contradiction. Hence, $s_i$ cannot intersect $s_j$
and the lemma follows. Below, we prove the above claim.

Suppose to the contrary there is a line $l$ contained in
$dw(s_i)\cap dw(s_j)$. Denote by $x_i$ (resp., $x_j$) the
intersection of the $x$-axis and $dw(s_i)$ (resp., $dw(s_j)$), e.g.,
see Fig.~\ref{fig:doublewedge}. Hence, both $x_i$ and $x_j$ are line
segments on the $x$-axis. Since the intersection of
$l(h_j)$ and the $x$-axis is strictly to the right of the intersection of
$l(h_{i+1})$ and the $x$-axis (since $H$ is $x$-intercept
ordered), $x_i$ does not intersect $x_j$. Since the slopes of both
$l(h_i)$ and $l(h_{i+1})$ are negative, a line contained in
$dw(s_i)$ must intersect $x_i$, and thus the line $l$ intersects $x_i$
(due to $l\subseteq dw(s_i)\cap dw(s_j)$).
Similarly, $l$ also intersects $x_j$. Hence, we obtain that $l$
intersects both $x_i$ and $x_j$. Since both $x_i$ and $x_j$ lie in
$x$-axis and $x_i$ does not intersect $x_j$,
the line $l$ has to be the $x$-axis. However, since the slopes of both
$l(h_i)$ and $l(h_{i+1})$ are negative, the
double wedge $dw(s_i)$ cannot contain the $x$-axis and thus cannot
contain $l$. Therefore, we obtain contradiction and the claim follows.
\end{myproof}

In light of Lemma \ref{lem:simple}, we can utilize the results in
\cite{ref:GuibasCo91}. Given a simple path in the plane,
{\em compact interval tree} data structures are proposed in
\cite{ref:GuibasCo91} to compute the convex hull of a query subpath
that is specified by the indices of the beginning vertex and the end
vertex of the subpath.
If applied to $\gamma$ in our problem, then after spending $O(n\log
n)$ time sorting the points in $H^*$ by their $x$-coordinates,
we have the following results: A data structure can be constructed in
$O(n\log\log n)$ time that can compute the lower hull $C^*_{ij}$ for
any query $q(i,j)$ in $O(\log n)$ time and $C^*_{ij}$ is represented
(by a compact interval tree) such that any standard convex hull
queries on $C^*_{ij}$ can be done in $O(\log n)$ time, where the
{\em standard convex hull queries} includes point-in-polygon tests,
finding intersections with lines, finding tangents through query
points, finding extreme vertices in a query direction, detecting
intersections of two polygons, and finding common tangents of
two convex hulls. Further, another data structure of construction time
$O(n)$ and query time $O(\log n\log\log n)$ is also given in
\cite{ref:GuibasCo91};  in
addition, there is also a data structure of construction time
$O(n\log^*n)$ and query time $O(\log n\log^* n)$ by making trade-off
between the construction and query \cite{ref:GuibasCo91}.
Both data structures are applicable to our problem.

By duality, the $x$-coordinate of each point $h_i^*\in H^*$
corresponds to the slope of the line $l(h_i)$. Thus, a sorted
order of the points in $H^*$ by $x$-coordinate corresponds to a
sorted order of the bounding lines of the half-planes in $H$ by
slope. The following lemma summarizes our discussions above.

\begin{lemma}\label{lem:summary}
In $O(n\log n)$ time, we can build a data structure that can answer
each 2-D sublist LP query in $O(\log n)$ time. Further, if the
bounding lines of the half-planes in $H$ are sorted by their slopes,
then there exist three data structures for the 2-D sublist LP
queries, whose construction time complexities are $O(n)$,
$O(n\log^*n)$, and $O(n\log\log n)$, respectively, and query time
complexities are $O(\log n\log\log n)$, $O(\log n\log^*n)$, and
$O(\log n)$, respectively.
\end{lemma}

According to the reduction procedure from the \problem\ problem to the
planar points
approximation problem \cite{ref:ChenEfkCenter11}, the slopes of the
bounding lines of the half-planes in $H$ correspond to the weights
of the points in the input point set $P$. Therefore, we have the following
corollary.

\begin{corollary}
There exists a data structure of time complexity $O(n\log n, \log
n)$ for the $\alpha(i,j)$ queries. Further, if the points of $\calP$
are sorted on the line $L$ and the weights of the points in $\calP$
are also sorted, then we have three data structures for the
$\alpha(i,j)$ queries of time complexities $O(n,\log n\log\log n)$,
$O(n\log^*n,\log n\log^*n)$, and $O(n\log\log n, \log n)$,
respectively.
\end{corollary}

By Lemma \ref{lem:100}, we have the following result for the \problem\
problem.

\begin{theorem}\label{theo:continuous}
The \problem\ problem is solvable in $O(n\log n)$ time. Further, if
the points of $\calP$ are sorted on $L$ and the weights of the
points in $\calP$ are also sorted, then the \problem\ is solvable in
$O(\min\{n+k^2\log^2{\frac{n}{k}}\log n\log\log
n,n\log^*n+k^2\log^2{\frac{n}{k}}\log n\log^*n,n\log\log
n+k^2\log^2{\frac{n}{k}}\log n\})$ time.
\end{theorem}

Therefore, if the points of $\calP$ are sorted on the line $L$ and
the weights of the points in $\calP$ are also sorted, for small $k$
(e.g., $k=O((\frac{n}{\log^3n\log\log n})^{1/2})$, which is true in
many applications), the \problem\
problem is solvable in $O(n)$ time.

\section{Conclusions}
\label{sec:conclusion}

In this paper, we give an $O(n\log n)$ time algorithm for the k-center
problem on a real line. In certain special cases, we can solve the
problem in linear time.
We also propose an efficient data structure to answer the 2-D sublist LP queries, which may find other applications.

As suggested by Tamir,
when $k=n-1$, the discrete unweighted \problem\ is equivalent to the Min
Gap problem, i.e., finding the closest pair of neighbors in $P$.
Hence, there is an $O(n\log n)$ time lower bound on the discrete
version. In fact, by the reduction from the Min Gap problem, we can
also show that the non-discrete unweighted \problem\ also has an
$O(n\log n)$ lower bound on the running time. Indeed, in any optimal
solution $OPT$ for $k=n-1$, there must be a
center at the middle position of the closest pair of neighbors in $P$,
and that center serves both neighbors; further, any
other center in $OPT$ serves one and only one demand point in $P$.
Therefore, given an optimal solution $OPT$, since the demand points
served by each center are known, the two demand points served by the
same center are the closest neighbors in $P$. We thus obtain the
$O(n\log n)$ time lower bound on the non-discrete unweighted \problem.

Note that the linear time algorithms for the unweighted
continuous/discrete k-center
problem on trees \cite{ref:FredericksonPa91} do not violate the
$O(n\log n)$ time lower
bound discussed above because the tree
structure already gives a partial order of the nodes in the tree.
An open problem is whether the techniques given in this paper can be
extended to the tree structure.

\section*{Acknowledgments}

We wish to thank Arie Tamir for his many helpful comments and suggestions.

%=====================end of %document=================

\footnotesize \baselineskip=11.0pt
\bibliographystyle{plain}
%\bibliography{reference}

%add appendix below
%\newpage
%\normalsize
%\appendix
%\section*{Appendix}

%\vspace{0.20in}
%\noindent
%{\bf Lemma \ref{lem:new10}}
%{\em A
%visibility tree $T_{vis}(\calS)$ of $\calS$ and $\calR$ can be
%computed in $O(n+h\log^{1+\epsilon}h)$ time.}
%\vspace{0.15in}

\end{document}